\documentclass[12pt]{cadabra}
\usepackage[colorlinks=false]{hyperref}

\begin{document}

\cdbmessagesOff

\title{\bfseries A brief introduction to Cadabra: a tool for tensor computations in General Relativity}
\author{%
Leo Brewin\\[10pt]%
School of Mathematical Sciences\\%
Monash University, 3800\\%
Australia}
\date{10-Mar-2009}
\reference{Preprint}

\maketitle

\begin{abstract}
\noindent
Cadabra is a powerful computer program for the manipulation of tensor equations. It was designed for use
in high energy physics but its rich structure and ease of use lends itself well to the routine
computations required in General Relativity. Here we will present a series of simple examples showing how
Cadabra may be used, including verifying that the Levi-Civita connection is a metric connection and a
derivation of the Gauss equation between induced and ambient curvatures.

\end{abstract}

\section{Computations in General Relativity}

Here are three examples of the kinds of computation that are often required in General Relativity.

\begin{description}
\item[Numerical computation.]\hfil\break\vskip-15pt
Use a numerical method to evolve the time symmetric initial data for a geodesic slicing of a
Schwarzschild spacetime in an isotropic gauge.

\item[Algebraic computation.]\hfil\break\vskip-15pt
Compute the Riemann tensor for the metric $ds^2=\Phi(r)^2\left(dr^2+r^2d\Omega^2\right)$.

\item[Tensor computations.]\hfil\break\vskip-15pt
Verify that $0=g_{ab;c}$ given
$\Gamma^{a}_{bc} = \frac{1}{2}g^{ad}\left(g_{dc,b} + g_{bd,c} - g_{bc,d}\right)$.
\end{description}

What tools might we use to perform these computations? For the first example, it is hard to envisage
\emph{not} using a computer to do the job. The second example is one which could easily be done by hand
or on a computer (using, for example, GRTensorII \cite{grtensorII:1994-01}). However, for the third
example, the vast majority of researchers in General Relativity would do the calculations by hand.
Recently a computer program, Cadabra \cite{peeters:2006-01,peeters:2007-02,peeters:2007-03}, has appeared
which may greatly assist us in these tedious tensor computations. This article will provide a brief
introduction to Cadabra and how it can be used in General Relativity.

There are a number of programs that, to varying degrees, can manipulate tensor expressions, including
GRTensorII \cite{grtensorII:1994-01},
MathTensor \cite{mathtensor:2008-01}, 
Canon \cite{manssur:2004-01}, 
Riemann \cite{portugal:1997-01}, 
xAct \cite{xact:2008-01}  
and my personal choice (and the subject of this article) Cadabra. It is not my intention to provide even a
cursory review of the existing computer tensor algebra systems. Instead I intend to show how Cadabra can
be used to do useful work in General Relativity. I suspect that all of the results reported in this
article can also be obtained using either MathTensor or xAct but as I have not done so I can not comment
on their relative merits.

I chose to work with Cadabra for a number of reasons, it uses LaTeX syntax for tensor expressions, it has
extensive and highly optimised routines for simplifying complex tensor expressions, it uses an
interpretive language that runs on top of a C++ program, it has a small but flexible grammar leading to
clear and readable code and it is in active development.

The use of (a subset of) LaTeX as the native language for Cadabra was the feature that first grabbed my
attention (apart from the promise of relieving me of the tedium of complex tensor gymnastics). You may
have already skimmed through this article and have wondered if I have massaged the Cadabra output for
inclusion in this article. The answer is an emphatic no. \emph{All of the output} appearing in this
article was included by a `cut-and-paste' operation directly from Cadabra's output. No stylistic changes
what so ever were made to the TeX produced by Cadabra. This ability to cut and paste between Cadabra
output and research documents is I believe one of Cadabra's unique strengths. This feature not only leads
to readable code (we only need to be literate in a simple subset of LaTeX) but it also minimises
transcription errors.


The following examples were deliberately constructed so as to require little mathematical development (for
the current audience) while being of sufficient complexity to allow Cadabra's features to be properly
showcased. I will make no assumptions about the dimensionality of the space, nor the signature of the
metric (other than that the metric is non-singular). However, I will assume that the connection is metric
compatible (i.e. the Levi-Civita connection). For the large part I will be using abstract index notation
but on the odd occasion where an explicit component based equation is given I will use a coordinate basis.
You might, as a simple exercise, like to adapt some of the following Cadabra examples to work with
non-coordinate basis (the changes are trivial).

\section{The Cadabra software}

The source code for Cadabra is freely available (but subject to the GPL licence) and has been successfully
compiled on many platforms including most flavors of Linux and MacOSX. It should compile on most standard
Unix like systems. The source code and binaries (for Linux) can be download from the web site
\cite{peeters:2007-03}. Cadabra for MacOSX is best installed (by my experience) from the source via the
MacPorts system \cite{macports:2008}.

Cadabra can be run either directly through the command line or through a purpose-built GUI (based on X11
and known as \verb|xcadabra|). To run cadabra over the file \verb|myinput.cdb| you need only type
\begin{cdbcode}
cadabra < myinput.cdb > myoutput.tex
\end{cdbcode}
from the command line. The \verb|xcadabra| GUI produces the same output but in a more convenient
environment. Further details can be found in the Cadabra documentation (available through the web site) 
\cite{peeters:2007-03}.

\section{A simple example}

How might we use Cadabra to verify that
$0=g_{ab;c}$ given $\Gamma^{a}_{bc} = \frac{1}{2}g^{ad}\left(g_{dc,b} + g_{bd,c} - g_{bc,d}\right)$?

This may seem an odd way to start but here is the full Cadabra code.
\vskip\parskip

\small
\Btag{01}
\begin{cdbcomment}
# --- The metric connection ---------------------------------------------------
\end{cdbcomment}
\begin{cdbverbatim}
{a,b,c,d,e,f,g,h,i,j,k,l,m,n,o,p,q,r,s,t,u#}::Indices.

g_{a b}::Metric.

\partial_{#}::PartialDerivative.

cderiv:=\partial_{c}{g_{a b}} - g_{a d}\Gamma^{d}_{b c}
                              - g_{d b}\Gamma^{d}_{a c};

Gamma:=\Gamma^{a}_{b c} -> (1/2) g^{a d} (  \partial_{b}{g_{d c}}
                                          + \partial_{c}{g_{b d}}
                                          - \partial_{d}{g_{b c}} );

@substitute!(cderiv)(@(Gamma));

@distribute!(
@eliminate_metric!(
@canonicalise!(
@collect_terms!(
\end{cdbverbatim}
\Etag{01}
\normalsize

The output from the above code is

\Btag{02}
\cdbeqtn{cderiv}%
{\partial_{c}{g_{a b}} - g_{a d} \Gamma^{d}\,_{b c} - g_{d b} \Gamma^{d}\,_{a c}}%
\cdbeqtn{Gamma}%
{\Gamma^{a}\,_{b c} \rightarrow \frac{1}{2}\, g^{a d} \left(\partial_{b}{g_{d c}} + \partial_{c}{g_{b d}} - \partial_{d}{g_{b c}}\right)}%
\cdbeqtn{cderiv}%
{\partial_{c}{g_{a b}} - \frac{1}{2}\, g_{a d} g^{d e} \left(\partial_{b}{g_{e c}} + \partial_{c}{g_{b e}} - \partial_{e}{g_{b c}}\right) - \frac{1}{2}\, g_{d b} g^{d e} \left(\partial_{a}{g_{e c}} + \partial_{c}{g_{a e}} - \partial_{e}{g_{a c}}\right)}%
\cdbeqtn{cderiv}%
{\partial_{c}{g_{a b}} - \frac{1}{2}\, g_{a d} g^{d e} \partial_{b}{g_{e c}} - \frac{1}{2}\, g_{a d} g^{d e} \partial_{c}{g_{b e}} + \frac{1}{2}\, g_{a d} g^{d e} \partial_{e}{g_{b c}} - \frac{1}{2}\, g_{d b} g^{d e} \partial_{a}{g_{e c}} - \frac{1}{2}\, g_{d b} g^{d e} \partial_{c}{g_{a e}} + \frac{1}{2}\, g_{d b} g^{d e} \partial_{e}{g_{a c}}}%
\cdbeqtn{cderiv}%
{\partial_{c}{g_{a b}} - \frac{1}{2}\, \partial_{b}{g_{a c}} - \frac{1}{2}\, \partial_{c}{g_{b a}} + \frac{1}{2}\, \partial_{a}{g_{b c}} - \frac{1}{2}\, \partial_{a}{g_{b c}} - \frac{1}{2}\, \partial_{c}{g_{a b}} + \frac{1}{2}\, \partial_{b}{g_{a c}}}%
\cdbeqtn{cderiv}%
{\partial_{c}{g_{a b}} - \frac{1}{2}\, \partial_{b}{g_{a c}} - \frac{1}{2}\, \partial_{c}{g_{a b}} + \frac{1}{2}\, \partial_{a}{g_{b c}} - \frac{1}{2}\, \partial_{a}{g_{b c}} - \frac{1}{2}\, \partial_{c}{g_{a b}} + \frac{1}{2}\, \partial_{b}{g_{a c}}}%
\cdbeqtn{cderiv}%
{0}%
\Etag{02}

Each of these line shows selected stages of processing by Cadabra. The zero in the final line is what
we were looking for -- it shows that $0=g_{ab;c}$.

We will now spend a moment of time to work through the above Cadabra code in some detail.

Comments in Cadabra are single lines that begin with the \verb|#| character in column one. Statements in
the Cadabra grammar fall into three distinct categories: \emph{properties}, \emph{expressions} and
\emph{algorithms}. The first three statements in the above code assign \emph{properties} to some symbols,
the next two statements define two \emph{expressions} with names \verb|cderiv| and \verb|Gamma| while the
remaining statements apply \emph{algorithms} to the expressions (i.e. they perform the computations). Note
that each of the statements ends with either a dot `\verb|.|' or a semi-colon `\verb|;|'. Only those
statements that finish with a semi-colon will produce output. If you look closely at the above you will
see that there are seven statements in the Cadabra code that end with a semi-colon and that there are
seven statements of Cadabra output. The statements and output are in exact one-to-one correspondence. You
can suppress the Cadabra output by terminating the statement with a colon `\verb|:|' rather than the
semi-colon `\verb|;|'.

What do these statements actually mean? The first statement 
\begin{cdbcode}
{a,b,c,d,e,f,g,h,i,j,k,l,m,n,o,p,q,r,s,t,u#}::Indices.
\end{cdbcode}
simply declares a set of symbols that may used as
indices. The last symbol \verb|u#| informs Cadabra that an infinite set of indices of the form 
\verb|u1,u2,u3|{\color{cdbverbatim}$\>\cdots$} is allowed. If you prefer to work with Greek indices then
you could declare
\begin{cdbcode}
{\alpha,\beta,\gamma,\mu,\nu,\theta,\phi#}::Indices.
\end{cdbcode}

The next statement
\begin{cdbcode}
g_{a b}::Metric.
\end{cdbcode}
declares the symbol \verb|g_{a b}| to represent a metric. This confers upon \verb|g_{a b}|
a raft of properties most notably that it is symmetric in its indices.

The third statement 
\begin{cdbcode}
\partial_{#}::PartialDerivative.
\end{cdbcode}
assigns to the symbol \verb|\partial| a derivative property. Note that the \verb|#| in
\verb|\partial_{#}| signifies that any number of indices are allowed. That is both \verb|\partial_{a}| and
\verb|\partial_{a b c d}| will be seen by Cadabra as derivative operators.

The next statement define an expression, \verb|cderiv|. 
\begin{cdbcode}
cderiv:=\partial_{c}{g_{a b}} - g_{a d}\Gamma^{d}_{b c} 
                              - g_{d b}\Gamma^{d}_{a c};

\end{cdbcode}
The name of the expression appears to the left of the `\verb|:=|' characters while the corresponding
tensor expression appears on the right using a familiar LaTeX syntax. Unlike LaTeX, we must always
separate the indices by one or more spaces. This ensures that Cadabra knows exactly how many indices
belong to an object (e.g.\ \verb|g_{ab}| would be interpreted as an object with \emph{one} covariant index
\verb|ab|).

Note carefully the braces around the metric term in \verb|\partial_{c}{g_{a b}}|. This is essential -- the
symbol \verb|\partial| is an operator and thus needs an argument to act on. And that argument is contained
inside the pair of braces.

How do we tell Cadabra that in the expression for \verb|cderiv| the symbols \verb|\Gamma^{a}{}_{bc}|
stands for the metric connection? That is, how do we couple the equation $\Gamma^{a}_{bc} =
\frac{1}{2}g^{ad}\left(g_{dc,b} + g_{bd,c} - g_{bc,d}\right)$ to the expression \verb|cderiv|? For this we
create two statements, the first is an expression that defines a substitution rule, the second uses
Cadabra's \verb|@substitute| algorithm to complete the job,

\begin{cdbcode}
Gamma:=\Gamma^{a}_{b c} -> (1/2) g^{a d} (  \partial_{b}{g_{d c}} 
                                          + \partial_{c}{g_{b d}} 
                                          - \partial_{d}{g_{b c}} );
@substitute!(cderiv)(@(Gamma));
\end{cdbcode}

The essence of this pair of statements is to substitute $\frac{1}{2}g^{ad}\left(g_{dc,b} + g_{bd,c} -
g_{bc,d}\right)$ wherever the symbol \verb|\Gamma^{a}_{b c}| appears in the expression \verb|cderiv|. This
may look simple but there are some important and subtle details that must be noted. The substitution rule
\verb|Gamma| as given above was for $\Gamma^{a}_{bc}$ yet in the expression for \verb|cderiv| we need
$\Gamma^{d}_{bc}$ and $\Gamma^{d}_{ac}$. Cadabra handles this index manipulation with ease, it will
relabel dummy indices in such a way as to avoid index clashes. This feature also exists in MathTensor and
xAct (but not so in GRTensorII). The construction \verb|@(...)| is Cadabra's way of referring to the
contents of an expression. The exclamation character `\verb|!|' in the statement tells Cadabra to impose
the substitution throughout the expression \verb|cderiv|. Without the `\verb|!|' character the
substitution will be applied once at most (i.e. on the first of the two $\Gamma^{a}_{bc}$'s in
\verb|cderiv|).

The remaining few statements

\begin{cdbcode}
@distribute!(
@eliminate_metric!(
@canonicalise!(
@collect_terms!(
\end{cdbcode}

serve only to massage the expression towards our expected result -- zero. The \verb|%| notation is Cadabra
shorthand for the last computed expression, in this case \verb|cderiv|. Each of the \verb|@...| statements
applies an algorithm to the expression \verb|cderiv|. The algorithm \verb|@distribute| is used to expand
products, it will expand \verb|a (b+c)| into \verb|a b + b c|. Earlier we gave \verb|g_{a b}| the property
\verb|::Metric|. This is used by the \verb|@eliminate_metric| algorithm to convert combinations such as
\verb|g^{a c} g_{c b}| into the Kronecker-delta $\delta^a{}_b$. The \verb|@canonicalise| algorithm is one
of Cadabra's most useful algorithms (on a par with \verb|@substitute|) as it can apply a wide range of
simplifications and general housekeeping on the expression. In this case it serves only to put the free
indices into ascending order (look carefully at the differences between the third and second last lines of
output). The final algorithm \verb|@collect_terms| needs no explanation.

\section{Covariant differentiation}

Cadabra does not have predefined algorithms for computing covariant derivatives, Riemann tensors, Ricci
tensor and so on. One of its strengths is that it provides a rich set of simple tools by which objects
such as those just noted can be constructed. As a second example we will now see how Cadabra can be
trained to compute covariant derivatives.

For a simple dual vector such as $v_a$ the textbook definition of the covariant derivative $v_{a;b}$ is
\[
v_{a;b} = v_{a,b} - \Gamma^{c}_{ab}v_{c}
\]
Here we will pursue a slight variation, one that lends itself well to computing higher order covariant
derivatives in Cadabra.

Choose any point $P$ in the spacetime and construct any curve through that point. Let $D^a$ be the
unit tangent vector to the curve, $A^a$ be a parallel vector field along the curve and let the curve
be parametrised by the proper distance $s$. Thus we have
\begin{spreadlines}{10pt}
\begin{gather*}
\frac{dv_a}{ds} = v_{a,b}D^b\\
0 = \nabla_D\>A^a = \frac{dA^a}{ds} + \Gamma^a_{bc}A^bD^c
\end{gather*}
\end{spreadlines}
Since $v_a A^a$ is a scalar function of $s$ we can compute its derivative along the curve in two ways, by
ordinary differentiation in $s$ or by applying the Leibniz rule to $\nabla_{D}\left(v_a A^a\right)$. This
leads to
\[
v_{a;b}A^a D^b = \frac{d(v_a A^a)}{ds}
\]
The left hand side is what we are looking for, while the right hand side is something we can easily train
Cadabra to perform. The final code contains two new algorithms, \verb|@prodsort| and \verb|@factor_out|.
Their actions are very simple, \verb|\@prodsort| rewrites any expression so that the symbols appear in a
particular order (usually in alphabetical order though this can be over-ridden with the
\verb|::SortOrder| property) and \verb|@factor_out| factors out nominated symbols from an expression. So,
without further ado here is the code.
\vskip\parskip

\small
\Btag{03}
\begin{cdbcomment}
# --- Covariant differentiation -----------------------------------------------
\end{cdbcomment}
\begin{cdbverbatim}
{a,b,c,d,e,f,g,h,i,j,k,l,m,n,o,p,q,r,s,t,u#}::Indices.

\partial_{#}::PartialDerivative.
\end{cdbverbatim}
\begin{cdbcomment}
# --- construct the scalar v_{a} A^{a}  ---------------------------------------
\end{cdbcomment}
\begin{cdbverbatim}
scalar:=v_{a} A^{a}:
\end{cdbverbatim}
\begin{cdbcomment}
# --- compute the derivative  -------------------------------------------------
\end{cdbcomment}
\begin{cdbverbatim}
derivD:=D^{c}\partial_{c}{@(scalar)}:
@distribute!(
@prodrule!(
@distribute!(
@substitute!(
@substitute!(
@prodsort!(
@rename_dummies!(
@canonicalise!(
\end{cdbverbatim}
\begin{cdbcomment}
# --- tidy up and display the results  ----------------------------------------
\end{cdbcomment}
\begin{cdbverbatim}
@factor_out!(scalar){A^{a}}:
@factor_out!(scalar){D^{a}};

@factor_out!(derivD){A^{a}}:
@factor_out!(derivD){D^{a}};
\end{cdbverbatim}
\Etag{03}
\normalsize

The above code produces the following output

\Btag{04}
\cdbmessage{@distribute}{not applicable.}
\cdbmessage{@substitute}{not applicable.}
\cdbmessage{@canonicalise}{not applicable.}
\cdbmessage{@factor_out}{not applicable.}
\cdbmessage{@factor_out}{not applicable.}
\cdbeqtn{scalar}%
{v_{a} A^{a}}%
\cdbeqtn{derivD}%
{A^{a} D^{b} \left(\partial_{b}{v_{a}} - \Gamma^{b}\,_{ac} v_{b}\right)}%
\Etag{04}

The right hand side is what we were seeking, $A^a D^b v_{a;b}$. As this quantity is also a scalar it
could be used as a starting point for further rounds of differentiation. Clearly this is easy to do and
would lead to expressions for higher order covariant derivatives of $v_a$. This is the basis for the
following example.

\section{Covariant differentiation and the Riemann tensor}

The somewhat unorthodox way in which we computed the covariant derivative in the previous example is
actually very well suited to our next example -- computing the Riemann tensor by commutation of
successive covariant derivatives. In this example we will `discover' that
\[
R_{a}{}^{d}{}_{bc} =   \partial_{b}{\Gamma^{d}{}_{a c}}
                     - \partial_{c}{\Gamma^{d}{}_{a b}} 
                     + \Gamma^{d}{}_{b e} \Gamma^{e}{}_{a c} 
                     - \Gamma^{d}{}_{c e} \Gamma^{e}{}_{a b}
\]
follows directly from $v_{a;b;c} - v_{a;c;b} = -R_{a}{}^{d}{}_{bc} v_{d}$

Here we will be using two covariant derivatives and thus, in the spirit of the previous example, we
imagine having two distinct curves through the point $P$. Let $D^a$ and $E^a$ be the respective unit
tangent vectors and let $u$ and $v$ be the respective proper distances along the curves (I could have
retained $s$ for the first curve but for aesthetics I prefer to use $u$ and $v$ here). This example is
slightly more complicated than the previous example but we still have considerable freedoms to choose the
nature of the vector fields $A^a$, $D^a$ and $E^a$ along the two curves through $P$. We will do the
calculations in two parts, first we will compute $d(d(v_aA^a)/du)dv$ then we reverse the order
$d(d(v_aA^a)/dv)du$. So for the first part we choose to parallel transport $A^a$ and $D^a$ along $E^a$. In
the second part we simply swap $D^a$ and $E^a$. Thus we have
\begin{spreadlines}{10pt}
\begin{gather*}
\frac{dv_a}{du} = v_{a,b}D^b\>,\quad\quad\quad
\frac{dv_a}{dv} = v_{a,b}E^b\\
0 = \nabla_D\>A^a = \frac{dA^a}{du} + \Gamma^a_{bc}A^bD^c\\
0 = \nabla_E\>A^a = \frac{dA^a}{dv} + \Gamma^a_{bc}A^bE^c\\
0 = \nabla_E\>D^a = \frac{dD^a}{dv} + \Gamma^a_{bc}D^bE^c\\
0 = \nabla_D\>E^a = \frac{dE^a}{dv} + \Gamma^a_{bc}E^bD^c
\end{gather*}
\end{spreadlines}
and this leads to
\[
\left(v_{a;b;c} - v_{a;c;b}\right)A^a D^b E^c 
= \left(\frac{d\ }{du}\frac{d\ }{dv} - \frac{d\ }{dv}\frac{d\ }{du}\right)(v_a A^a)
\]
The right hand side is very easy to implement in Cadabra -- just two (almost identical) rounds of
the code given in the previous example. Here is the Cadabra code.
\vskip\parskip

\small
\Btag{05}
\begin{cdbcomment}
# --- covariant differentiation and the Riemann tensor ------------------------
\end{cdbcomment}
\begin{cdbverbatim}
{a,b,c,d,e,f,g,h,i,j,k,l,m,n,o,p,q,r,s,t,u#}::Indices.

\partial_{#}::PartialDerivative.
\end{cdbverbatim}
\begin{cdbcomment}
# --- force Gamma to be symmetric in its lower two indices --------------------
\end{cdbcomment}
\begin{cdbverbatim}
\Gamma^{a}_{b c}::TableauSymmetry(shape={2}, indices={1,2}).
\end{cdbverbatim}
\begin{cdbcomment}
# --- construct the scalar v_{a} A^{a}  ---------------------------------------
\end{cdbcomment}
\begin{cdbverbatim}
scalar:=v_{a} A^{a}:
\end{cdbverbatim}
\begin{cdbcomment}
# --- compute the covariant derivative in the direction of D^a  ---------------
\end{cdbcomment}
\begin{cdbverbatim}
derivD:=D^{c}\partial_{c}{@(scalar)}:
@distribute!(
@prodrule!(
@distribute!(
@substitute!(
@prodsort!(
@rename_dummies!(
@canonicalise!(
\end{cdbverbatim}
\begin{cdbcomment}
# --- compute the covariant derivative in the direction of E^a  ---------------
\end{cdbcomment}
\begin{cdbverbatim}
derivDE:=E^{c}\partial_{c}{@(derivD)}:
@distribute!(
@prodrule!(
@distribute!(
@substitute!(
@substitute!(
@prodsort!(
@rename_dummies!(
@canonicalise!(
\end{cdbverbatim}
\begin{cdbcomment}
# --- copy to derivED then swap the order of the derivatives ------------------
\end{cdbcomment}
\begin{cdbverbatim}
derivED:=@(derivDE):

@substitute!(
@substitute!(
@substitute!(
\end{cdbverbatim}
\begin{cdbcomment}
# --- compute difference in mixed covariant derivatives -----------------------
\end{cdbcomment}
\begin{cdbverbatim}
diff:=@(derivDE) - @(derivED):
\end{cdbverbatim}
\begin{cdbcomment}
# --- tidy up and display the results  ----------------------------------------
\end{cdbcomment}
\begin{cdbverbatim}
{A^{a},D^{a},E^{a},v_{a},\Gamma^{a}_{b c}}::SortOrder.

@prodsort!(
@rename_dummies!(
@canonicalise!(
@collect_terms!(

@factor_out!(
@factor_out!(
@factor_out!(
@factor_out!(

@print["A^a D^b E^c (v_{a;b;c}-v_{a;c;b})="~@(diff)];
\end{cdbverbatim}
\Etag{05}
\normalsize

Here is the output form the above code.

\Btag{06}
\cdbmessage{@distribute}{not applicable.}
\cdbmessage{@canonicalise}{not applicable.}
\cdbeqtn{derivDE}%
{- A^{a} D^{b} E^{c} \Gamma^{d}\,_{a c} \partial_{b}{v_{d}} - A^{a} D^{b} E^{c} \Gamma^{d}\,_{b c} \partial_{d}{v_{a}} + A^{a} D^{b} E^{c} \partial_{b c}{v_{a}} + A^{a} D^{b} E^{c} \Gamma^{d}\,_{a c} \Gamma^{e}\,_{b d} v_{e} + A^{a} D^{b} E^{c} \Gamma^{d}\,_{a e} \Gamma^{e}\,_{b c} v_{d} - A^{a} D^{b} E^{c} \partial_{c}{\Gamma^{d}\,_{a b}} v_{d} - A^{a} D^{b} E^{c} \Gamma^{d}\,_{a b} \partial_{c}{v_{d}}}%
\cdbeqtn{derivED}%
{- A^{a} E^{b} D^{c} \Gamma^{d}\,_{a c} \partial_{b}{v_{d}} - A^{a} E^{b} D^{c} \Gamma^{d}\,_{b c} \partial_{d}{v_{a}} + A^{a} E^{b} D^{c} \partial_{b c}{v_{a}} + A^{a} E^{b} D^{c} \Gamma^{d}\,_{a c} \Gamma^{e}\,_{b d} v_{e} + A^{a} E^{b} D^{c} \Gamma^{d}\,_{a e} \Gamma^{e}\,_{b c} v_{d} - A^{a} E^{b} D^{c} \partial_{c}{\Gamma^{d}\,_{a b}} v_{d} - A^{a} E^{b} D^{c} \Gamma^{d}\,_{a b} \partial_{c}{v_{d}}}%
\cdbeqtn{diff}%
{A^{a} D^{b} E^{c} v_{d} \left(\Gamma^{d}\,_{b e} \Gamma^{e}\,_{a c} - \partial_{c}{\Gamma^{d}\,_{a b}} - \Gamma^{d}\,_{c e} \Gamma^{e}\,_{a b} + \partial_{b}{\Gamma^{d}\,_{a c}}\right)}%
\cdbprint%
{A^a D^b E^c \left(v_{a;b;c}-v_{a;c;b}\right)=A^{a} D^{b} E^{c} v_{d} \left(\Gamma^{d}\,_{b e} \Gamma^{e}\,_{a c} - \partial_{c}{\Gamma^{d}\,_{a b}} - \Gamma^{d}\,_{c e} \Gamma^{e}\,_{a b} + \partial_{b}{\Gamma^{d}\,_{a c}}\right)}%
\Etag{06}

The last line of output is what we have been seeking and apart from it being mathematically correct it is
worth noting how that line was created. The right hand side was obtained from a result computed by
Cadabra while the left hand side was a string supplied by us. This construction is performed in the
\verb|@print[...]| algorithm. You can see it in the last line of the Cadabra code. Its use is clear --
it creates nicely formatted output. But there is a potential danger here -- the onus falls on the
user (us) to ensure that the left hand side is what it should be, Cadabra does no processing what so ever
on the argument to \verb|@print|. So if you chose to write nonsense such as

\begin{cdbcode}
@print["A_a + B_c = g_{ab}"];
\end{cdbcode}

then Cadabra will dutifully obey.

The output for this example could have been restricted to just the final line (by changing the appropriate
`\verb|;|' to `\verb|:|'). But there is value in including output from other lines for it shows
how the computations unfold and it allows us to verify that Cadabra is doing what we think and want
it to do. Output like this is very useful when first writing and later debugging Cadabra code.

There is one very important part of the calculations that we have not yet spoken about -- How do we
tell Cadabra that we are using a torsion free connection? The line in the above code that does the
job is
\begin{cdbcode}
\Gamma^{a}_{b c}::TableauSymmetry(shape={2}, indices={1,2}).
\end{cdbcode}
This tells Cadabra that we want the
connection to be symmetric in its lower two indices (i.e. torsion free). 

Cadabra uses sophisticated algorithms to handle tensor symmetries based on the Littlewood-Richardson
algorithm for finding a basis of the irreducible representations of totally symmetric groups. The
algorithm uses Young diagrams which consist of a set of cells arranged as series of rows which in turn are
described by the \verb|::TableauSymmetry| property. In short, the index symmetries of a tensor are encoded
in these diagrams. The \verb|shape={...}| parameter describes the shape of a Young diagram, in this case
it consists of one row with two cells. The \verb|indices={...}| parameter describes how the tensor's
indices are assigned to the cells. For this purpose, the indices on the tensor are counted from left to
right starting with zero. So in the above example the lower two indices $b$ and $c$ are counted as 1 and 2
and they are assigned to the two cells of the Young diagram. More details on using tableaux as a way to
describe tensor symmetries can be found in the Cadabra manual.

If Young diagrams and tableaux are not your cup of tea then there is a (less than ideal) alternative.
One way to obtain a symmetric connection is to temporarily put $\Gamma^{a}_{bc} = G^{a} G_{bc}$
where $G_{bc}=G_{cb}$, ask Cadabra to make its simplifications and then return the $\Gamma^{a}_{bc}$
to the result. This is not a mathematical operation, it is just a trick to help Cadabra spot what
symmetries are available. Here is a fragment of code that does the job (assuming \verb|%| is an
expression that contains $\Gamma^{a}_{bc}$)
\vskip\parskip

\small
\Btag{20}
\begin{cdbcomment}
# --- trick to impose zero torsion (symmetric connection) ---------------------
\end{cdbcomment}
\begin{cdbverbatim}
G_{a b}::Symmetric.

@substitute!(
@prodsort!(
@canonicalise!(
@collect_terms!(
@substitute!!(
\end{cdbverbatim}
\Etag{20}
\normalsize 

The problem with this approach is that if the pair of terms $G^{a}$ and $G_{bc}$ ever get
separated (e.g. from a product rule) then it may not be possible to complete the last step
of this trick, that is, to eliminate the $G^{a}$ and $G_{ab}$ in favour of $\Gamma^{a}_{bc}$.
If on the other hand you can be sure that such problems can not arise (e.g. you apply the trick after
all the derivatives have been computed) then this method is rather easy to apply. It also
provides a quick way to implement more complicated symmetries (e.g. if $A_{abcde}$ is symmetric
in the first two and last three indices put $A_{abcde} = G_{ab}G_{cde}$).

\section{The Gauss equation}

Lest it be thought that every covariant derivative in Cadabra needs to be cast in the form given in the
previous examples, here is an application of Cadabra to the derivation of the Gauss equation relating the
induced and ambient curvatures of a hypersurface in an $n-$dimensional Riemannian manifold.

We shall start with a brief review of the underlying mathematics. Suppose $\Sigma$ is an
$(n-1)-$dimensional subspace of an $n-$dimensional space $M$. Suppose $M$ is equipped with Riemannian
metric $g$ and a metric compatible derivative operator $\nabla$. The subspace $\Sigma$ will, by way of its
embedding in $M$, inherit a metric and derivative operator which we will denote by $h$ and $D$
respectively (note this use of $D$ differs from that in the previous sections, here $D$ is a differential
operator). Let $n^a$ be the oriented unit normal to $\Sigma$. Then the metrics of $\Sigma$ and $M$ are
related by
\[
g_{a b} = h_{a b} + n_{a} n_{b}
\]
while, for any dual-vector $v_a$ lying in $\Sigma$ (i.e. $v_a n^a=0$), we have
\[
D_{b} v_{a} = h^{d}{}_{b} h^{c}{}_{a} \> \nabla_{d} v_{c}
\]
where $h^{a}{}_{b} = g^{a}{}_{b} - n^{a} n_{b}$ is the projection operator. The curvature tensor for
$(\Sigma,h,D)$ can then be obtained by computing
$\left(D_{c} D_{b}-D_{b} D_{c}\right) v_{a}$.
This is all very standard and can be found in most textbooks on differential geometry
(see \cite{chavel:2006-01}).

Translating these equations into Cadabra code is very straightforward and follows a now familiar pattern.
Unlike the previous examples, we will begin by discussing fragments of code that express the basic
mathematical relations as just given. These code fragments will later be glued together to form a complete
Cadabra program.

We will start with the definition of the projection operator $h^{a}{}_{b} = g^{a}{}_{b}-n^{a}n_{b}$ and
its use in defining $D$ in terms of $\nabla$. We will use the symbol \verb|hab| to record the projection
operator and \verb|vpq| to record the covariant derivative $D_{q}v_p$. Thus our code will contain the
lines
\begin{cdbcode}
hab:=h^{a}_{b} -> g^{a}_{b} - n^{a} n_{b}:
vpq:=v_{p q} -> h^{a}_{p}h^{b}_{q}\nabla_{b}{v_{a}}:
\end{cdbcode}
We will also need an expression for the commutation of the covariant derivatives,\hfill\break
$\left(D_{r}D_{q} - D_{q}D_{r} \right)v_{p}$ which we write as \verb|vpqr|
\begin{cdbcode}
vpqr:=h^{a}_{p}h^{b}_{q}h^{c}_{r} ( \nabla_{c}{v_{a b}} - \nabla_{b}{v_{a c}} ):
@substitute!(vpq)(@(hab)):
@substitute!(vpqr)(@(vpq)):
\end{cdbcode}

Finally we will need to introduce some standard substitutions to simplify and tidy the result. Note that
\emph{all of the previous definitions and following substitutions are exactly what we would normally do if
we were to do these calculations by hand}. For example, the lines
\begin{cdbcode}
@substitute!(
@substitute!(
\end{cdbcode}
expresses the condition that $n^a$ is normal to the subspace, $0 = n^b h^{a}{}_{b}$ and
$0 = n^{b}h_{b}{}^{a}$. The line
\begin{cdbcode}
@substitute!(
\end{cdbcode}
states that the covariant derivative of $g$ is zero while the line 
\begin{cdbcode}
@substitute!(
\end{cdbcode}
is a simple re-working of $0=\nabla\left(n^a v_a\right)=\left(\nabla n^a \right)v_a + n^a
\left(\nabla v_a\right)$ to eliminate first derivatives of $v^a$ from the expression \verb|vpqr|. The
next line
\begin{cdbcode}
@substitute!(
\end{cdbcode}
squeezes a projection operator between $v_{a}$ and $\nabla n^a$. This is allowed because $v^a$ has zero
normal component. Finally, lines like
\begin{cdbcode}
@substitute!(
@substitute!(
\end{cdbcode}
can be used to introduce the extrinsic curvature tensor $K_{ab}$.

Clearly we have to supplement the above code fragments with extra statements, such as an index set,
substitution and simplification rules etc., before Cadabra can do its job. Such pieces of code are very
similar to those given in the previous examples and thus require no further explanation here. Here then is
the final code.
\vskip\parskip

\small
\Btag{07}
\begin{cdbcomment}
# --- The Gauss equation ------------------------------------------------------
\end{cdbcomment}
\begin{cdbverbatim}
::PostDefaultRules( @@collect_terms!(
{a,b,c,d,e,f,g,i,j,k,l,m,n,o,p,q,r,s,t,u#}::Indices(position=fixed).

\nabla_{#}::Derivative.

K_{a b}::Symmetric.
g^{a}_{b}::KroneckerDelta.
\end{cdbverbatim}
\begin{cdbcomment}
# --- define the projection operator ------------------------------------------
\end{cdbcomment}
\begin{cdbverbatim}
hab:=h^{a}_{b} -> g^{a}_{b} - n^{a} n_{b}:
\end{cdbverbatim}
\begin{cdbcomment}
# --- 3-covariant derivative obtained by projection on 4-covariant derivative -
\end{cdbcomment}
\begin{cdbverbatim}
vpq:=v_{p q} -> h^{a}_{p}h^{b}_{q}\nabla_{b}{v_{a}}:
\end{cdbverbatim}
\begin{cdbcomment}
# --- compute 3-curvature by commutation of covariant derivatives -------------
\end{cdbcomment}
\begin{cdbverbatim}
vpqr:=h^{a}_{p}h^{b}_{q}h^{c}_{r} ( \nabla_{c}{v_{a b}} - \nabla_{b}{v_{a c}} ):

@substitute!(vpq)(@(hab)):
@substitute!(vpqr)(@(vpq)):

@distribute!(
@prodrule!(
@distribute!(
@eliminate_kr!(
\end{cdbverbatim}
\begin{cdbcomment}
# --- standard substitutions --------------------------------------------------
\end{cdbcomment}
\begin{cdbverbatim}
@substitute!(
@substitute!(
@substitute!(
@substitute!(
@substitute!(
@substitute!(
@substitute!(
\end{cdbverbatim}
\begin{cdbcomment}
# --- tidy up and display the results -----------------------------------------
\end{cdbcomment}
\begin{cdbverbatim}
{h^{a}_{b},\nabla_{a}{v_{b}}}::SortOrder.

@prodsort!(
@rename_dummies!(
@canonicalise!(
@factor_out!!(
\end{cdbverbatim}
\Etag{07}
\normalsize

The output returned by Cadabra is

\Btag{08}
\cdbmessage{@substitute}{not applicable.}
\cdbeqtn{vpqr}%
{h^{a}\,_{p} h^{b}\,_{q} h^{c}\,_{r} \left(\nabla_{c}{\nabla_{b}{v_{a}}} - \nabla_{b}{\nabla_{c}{v_{a}}}\right) + K_{p r} K_{q}\,^{a} v_{a} - K_{p q} K_{r}\,^{a} v_{a}}%
\Etag{08}

which, although correct, is not in the form we are most familiar with. We could reformat the output by
including suitable \verb|@print| statements or we could wrest control from Cadabra and do the final
tidying up by hand (we are not invalids and it's wise to retain our skills).

Recall that \verb|vpqr| was defined to be $\left(D_{r}D_{q} - D_{q}D_{r}\right)v_{p}$. 
If we use this and the Ricci identity for curvatures $v_{a;b;c}-v_{a;c;b} =
-R_{a}{}^{d}{}_{bc}v_d$ we very easily recover the Gauss equation in (one of) its familiar forms
\[
\left({}^hR_{p}{}^{s}{}_{qr}\right) 
   = h^{a}{}_{p}h^{s}{}_{d}h^{b}{}_{q}h^{c}{}_{r}\left({}^g R_{a}{}^{d}{}_{bc}\right)
   + K_{pr}K_{q}{}^{s} - K_{pq}K_{r}{}^{s}
\]

\section{Outlook}

Simple problems such as those given here can be done easily by hand (or passed off to a graduate student).
But they tell us nothing about Cadabra's utility or performance for large scale tensor computations in
General Relativity. In a follow-up paper \cite{brewin:2009-03} we will report on the use of Cadabra for
higher order Riemann normal expansions of the metric, connection and solutions to the initial and boundary
value problems for the geodesic equations. The results have been completed up to 6th order in the
curvature tensor.

It is my opinion that Cadabra is very well suited to large scale tensor algebra computations
in General Relativity and will be a welcome addition to the relativist's computational toolkit.

\section{Source}

A .tar.gz archive of the Cadabra files used in preparing this paper can be found at this
URL \url{http://users.monash.edu.au/~leo/research/papers/files/lcb09-02.html}

\section{Acknowledgements}

I am very grateful to Kasper Peeters for his many helpful suggestions. Any errors,
omissions or inaccuracies in regard to Cadabra are entirely my fault (I hope there are none).


\providecommand{\href}[2]{#2}\begingroup\raggedright\endgroup

\end{document}